\documentclass[elsart.sty]{elsart} 
\usepackage{epsfig}
\usepackage{xspace}
\begin{document}
\newlength{\caheight}
\setlength{\caheight}{12pt}
\multiply\caheight by 7
\newlength{\secondpar}
\setlength{\secondpar}{\hsize}
\divide\secondpar by 3
\newlength{\firstpar}
\setlength{\firstpar}{\secondpar}
\multiply\firstpar by 2
\def\bce{\begin{center}}  \def\ece{\end{center}}
\def\bit{\begin{itemize}}    \def\eit{\end{itemize}}
\def\ben{\begin{enumerate}}    \def\een{\end{enumerate}}
\newcommand\cpc[3]{{Comput. Phys. Commun. }{\bf #1} (#2) #3}
\newcommand\npb[3]{{Nucl. Phys. }{\bf B #1} (#2) #3}
\newcommand{\hepph}[1]{{\tt hep-ph/#1}}
\newcommand\plb[3]{{Phys. Lett. }{\bf B #1} (#2) #3}
\newcommand\prd[3]{{Phys. Rev. }{\bf D #1} (#2) #3}
\newcommand\sjnp[3]{{Sov. J. Nucl. Phys. }{\bf #1} (#2) #3}
\newcommand\jetp[3]{{Sov. Phys. JETP }{\bf #1} (#2) #3}
\newcommand\zpc[3]{{Z. Phys. }{\bf C #1} (#2) #3}
\newcommand\ibid[3]{{\it ibid.}{\bf #1} (#2) #3}
\newcommand\jhep[3]{{J. High Energy Phys. }{\bf #1} (#2) #3}
\newcommand\ptp[3]{{Prog. Theor. Phys. }{\bf #1} (#2) #3}
\hyphenation{BA-BA-YA-GA}
\hfill
\parbox[0pt][\caheight][t]{\secondpar}{
  \rightline
  {\tt \shortstack[l]{
      FNT/T-2000/05\\
      SISSA/28/2000/EP}}
  }
\begin{frontmatter}
\title{Large-angle Bhabha scattering and 
luminosity at flavour factories}
\author[Pavia1]{C.M.~Carloni Calame}
\author[Trieste]{C.~Lunardini}
\author[Pavia1]{G.~Montagna}
\author[Pavia2]{O.~Nicrosini} 
\author[Pavia2]{F.~Piccinini}
\address[Pavia1]{Dipartimento di Fisica Nucleare e Teorica,
Universit\`a di Pavia, 
and INFN, Sezione di Pavia, 
via A.Bassi 6, 27100, Pavia, Italy}
\address[Trieste]{SISSA-ISAS, via Beirut 2-4, 34100, Trieste, Italy and 
INFN, Sezione di Trieste, via Valerio 2, 34127, Trieste, Italy}
\address[Pavia2]{INFN, Sezione di Pavia, and 
Dipartimento di Fisica Nucleare e Teorica,
Universit\`a di Pavia, 
via A.~Bassi 6, 27100, Pavia, Italy}
\begin{abstract}
The luminosity determination of electron-positron colliders 
operating in the region of low-lying hadronic resonances 
($E_{cm} \simeq$ 1-10~GeV), such as BEPC/BES, DA$\Phi$NE, 
KEKB, PEP-II and VEPP-2M, requires the precision 
calculation of the Bhabha process at large scattering angles.
In order to achieve a theoretical accuracy at a few 0.1\% 
level, the inclusion of radiative corrections is mandatory. 
The phenomenologically relevant effect of QED corrections 
is taken into account in the framework of the Parton Shower 
(PS) method, which is employed both for cross section calculation 
and event generation. To test the reliability of the 
approach, a benchmark calculation, including 
exact $O (\alpha)$ corrections and higher-order 
leading logarithmic contributions, is developed as well 
and compared in detail with the PS predictions. 
The effect of $O(\alpha)$ next-to-leading and 
higher-order leading corrections is investigated in the presence 
of realistic event selections for the Bhabha process 
at the $\Phi$ factories. A new Monte Carlo generator 
 for data analysis (BABAYAGA) is presented, with an estimated 
accuracy of 0.5\%. Possible developments aiming at 
improving its precision 
and range of applicability are discussed.
\end{abstract}
\end{frontmatter}

\section{Introduction}
The accurate determination of the machine luminosity 
is a fundamental ingredient for the successful accomplishment
of the physics programme of electron-positron ($e^+ e^-$) colliders 
operating in the region of the low-lying hadronic 
resonances, such as BEPC/BES (Beijing), DA$\Phi$NE 
(Frascati), VEPP-2M (Novosibirsk), as well as for 
the BELLE and BABAR experiments around the $\Upsilon$ at 
KEKB (KEK) and PEP-II (SLAC). 
In particular, the precise measurement of the 
hadronic cross section at the $\Phi$ factories requires a 
luminosity determination with a total relative error 
better than 1\%~\cite{dh1,dh2,gp,cv}. As well known, the luminosity $L$ 
of $e^+ e^-$ colliders can be precisely derived via the relation
$L = N / \sigma_{th}$, where $N$ and $\sigma_{th}$ are the number of events
and the theoretical cross section 
of a given reference reaction, respectively. In order to make 
the total (experimental and theoretical) luminosity
error as small as possible, 
the cross section $\sigma_{th}$ of the reference process 
should be large, in order 
to keep the statistical uncertainty small, and  
calculable with high theoretical accuracy.

At $e^+ e^-$ machines operating in the energy range 1-10~GeV, 
the best candidate 
fulfilling the above criteria is the process 
$e^+ e^-  \to e^+ e^-$ (Bhabha scattering) detected at 
large scattering angles, say in the 
angular range $20^\circ \leq \vartheta \leq 160^\circ$. 
For example, at the $\Phi$ factory DA$\Phi$NE the KLOE detector can be used 
to detect such events~\cite{cv}, where the Bhabha scattering 
cross section is significant, being of the order of 
$10^4$~nb at a center of mass (c.m.) energy 
around the $\Phi$ resonance ($\sqrt{s} \simeq 1$~GeV). 

Therefore, on theoretical side, precision calculations of the 
large-angle Bhabha (LABH) cross section are 
demanded, with a theoretical accuracy at a few 0.1\%
level. This implies to include in the calculation  
all the phenomenologically relevant radiative 
corrections, in particular the large effects due to 
photonic radiation. Furthermore, such effects should be
implemented and accurately simulated 
in event generators, which are 
strongly demanded by the experimental analysis. 

At present, the status of the 
theoretical predictions and 
generators of interest for the LABH process 
at low-energy $e^+ e^-$ machines can be summarized as follows.
An exact $O(\alpha)$ generator, based on the 
calculation of ref.~\cite{bmc} and modified to 
match DA$\Phi$NE characteristics, is used in Monte Carlo 
(MC) studies by the KLOE collaboration~\cite{cv,dv}. An independent
$O(\alpha)$ generator is also in use by the CMD-2 and
SND experiments at VEPP-2M~\cite{priv}. In both the programs used in 
such MC simulations the effect of higher-order 
corrections is not taken into account. A semi-analytical calculation 
of the cross section for large-angle QED processes, i.e. 
$e^+ e^- \to \mu^+\mu^-, e^+e^-,
\gamma\gamma$, below 3 GeV was performed in ref.~\cite{abaetal}. 
It includes exact $O(\alpha)$ plus leading logarithmic 
(LL) higher-order corrections. 
This formulation is available as a computer code described in 
ref.~\cite{labsmc}.
The recently developed Bhabha generator BHWIDE~\cite{bhwide}, 
based on the Yennie-Frautschi-Suura approach for the treatment of QED 
radiation, 
appears in the list of simulation tools presently under 
consideration by the BABAR collaboration at the $B$ 
factory PEP-II~\cite{babar}. A QED Parton Shower (PS) algorithm, 
which is employed in the present study, is adopted in 
refs.~\cite{fuetal,unibab} for the computation 
of radiative corrections to LABH scattering. These 
calculations, however, are optimized to high-energy LABH 
 and differ, as it will be discussed, in some aspects from the 
present implementation of the PS model. 
A complete inventory of existing calculations and 
programs, for both small- and large-angle Bhabha 
scattering, used at very high-energy $e^+ e^-$ colliders 
can be found in ref.~\cite{lep2bha}.

The paper is organized as follows. 
In Sect.~2 the theoretical formulation,  
which is based upon a QED realization of the PS method 
to account for radiative corrections due to photon emission, 
is described. The steps and kinematics of the algorithm are
reviewed. Sect.~3 is devoted 
to the description of a new, original PS 
generator for the simulation of the LABH process and based upon 
the formulation previously discussed. A first
sample of numerical results from the PS Bhabha generator
is also given, with particular emphasis on the simulation of the Bhabha 
process at the $\Phi$ factories in the 
presence of realistic event selections (ES). In the following Sections tests
of the reliability of the PS approach are shown and commented,
both at the level of integrated cross sections and 
differential distributions. To this end, the calculation 
of the exact $O(\alpha)$ cross section and 
its matching with higher-order LL
corrections is addressed in Sect.~4. This is meant as 
a benchmark calculation, developed in order to 
check the physical+technical precision of the PS
generator. Detailed comparisons between the PS predictions and the results of
the benchmark computation are given in Sect.~5. Conclusions, open issues and
possible developments are discussed in Sect.~6.        

\section{Theoretical approach and the Parton Shower method}
In order to approach the aimed theoretical accuracy, 
the calculation of the QED corrected 
Bhabha scattering cross section and the 
relative event generation is performed according to the 
master formula~\cite{prd}:
\begin{eqnarray}
&\sigma(s)=\int dx_- dx_+ dy_- dy_+ \int d\Omega_{lab}
D(x_-,Q^2)D(x_+,Q^2) \times & \nonumber\\ 
&D(y_-,Q^2)D(y_+,Q^2) \frac{d\sigma_0}{d\Omega_{cm}}
\big(x_-x_+s,\vartheta_{cm}\big)
J\big(x_-,x_+,\vartheta_{lab}\big)\Theta(cuts) , & 
\label{eq:sezfs}
\end{eqnarray}   
which is based on the factorization theorems 
of (universal) infrared and 
collinear singularities. Equation~(\ref{eq:sezfs})
can be worked out within a QED PS algorithm for the calculation 
of the electron Structure Function (SF) $D(x,Q^2)$, both for 
initial-state radiation (ISR) and final-state radiation (FSR). In 
eq.~(\ref{eq:sezfs}) 
$d\sigma_0 / d\Omega$ is the Born-like differential 
cross section relevant for centre c.m. 
energy between 1-10~GeV, 
including the photonic $s$- and $t$-channel
diagrams and their interference,  
and the contributions due to exchange of 
vector resonances, such as $\Phi$, 
$J/\Psi$ and $\Upsilon$. Following the standard procedure 
described in ref.~\cite{gmnp}, the contribution of hadronic resonances
is taken into account in terms of their effective couplings 
to the electron. At c.m. energy around 1 GeV, 
the total $\Phi$ contribution 
amounts to $\approx 0.1(0.3)\%$ for 
$20^\circ(50^\circ) \leq \vartheta \leq 160^\circ(130^\circ)$, 
where $\vartheta$ is the electron scattering angle. For higher 
energies, as in the case of BELLE and BABAR experiments around 
the $\Upsilon$, the effect of the $\Upsilon$ resonance to 
the Bhabha cross section is of the same order. 
It is worth noticing that for the low-energy colliders
the LABH cross section is largely dominated by $t$-channel 
photon exchange; hence, its leading dynamics is quite similar 
to small-angle Bhabha (SABH) at LEP1/SLC and LABH at LEP2 and 
higher energies~\cite{lep2bha}. 
In the hard-scattering 
cross section, the relevant correction due to  
vacuum polarization is taken into account as well, 
by adopting the parameterization of ref.~\cite{vpol}.  
In the evaluation of the hadronic contribution 
to the vacuum polarization, the Euclidean 
value of the momentum in the photon propagator 
has been used as the appropriate scale for 
time-like momenta~\cite{vpol1}. The effect of the 
running coupling constant at $\sqrt{s} = M_\Phi$ is to enhance 
the cross section by 
$\approx 2(2.5)\%$ for $20^\circ(50^\circ) \leq 
\vartheta \leq 160^\circ(130^\circ)$. The factor 
$J\big(x_-,x_+,\vartheta_{lab}\big)$ in eq.~(\ref{eq:sezfs}) 
accounts for the
boost from the c.m. to the lab frame due to emission of unbalanced 
ISR from the
electron and positron legs, while $\Theta(cuts)$ 
represents cuts implementation. 

The basic ingredient of eq.~(\ref{eq:sezfs}) is the electron SF $D(x,Q^2)$, 
which represents the probability density 
of finding ``inside'' a parent electron an electron 
with momentum fraction $x$ and virtuality $Q^2$. It 
can be explicitly obtained in QED by solving 
the Dokshitzer-Gribov-Lipatov-Altarelli-Parisi (DGLAP) evolution 
equation~\cite{ap} in the non-singlet channel:
\begin{equation}  
Q^2\frac{\partial}{\partial Q^2}D(x,Q^2)=
\frac{\alpha}{2\pi}\int_x^{1}\frac{dy}{y}P_+(y) D(\frac{x}{y},Q^2)  ,  
\label{eq:ap} 
\end{equation}
where $P_+(x)$ is the regularized $e \to e + \gamma$ 
splitting function 
\begin{equation} 
P_+(x)=\frac{1+x^2}{1-x}-\delta(1-x)\int_0^1 dt P(t) .
\label{eq:vertex}
\end{equation} 
Notice that the energy scale $Q^2$ entering the 
SF will be, in general, 
dependent on the specific process under study. 
No {\em exact} analytical 
solution of eq.~(\ref{eq:ap}) is known in the literature. 
After the efforts undertaken 
in the last decade for the program of precision physics at
LEP/SLC, the theoretical situation can be 
summarized as follows (see for instance ref.~\cite{rnc} 
and references therein):
\ben
\item approximate (accurate) analytical solutions 
in the collinear limit~\cite{sfcoll}. 
This kind of SFs is implemented in most of the 
programs for data analysis at LEP/SLC~\cite{rnc};
\item exact numerical solution, obtained via a numerical 
calculation of the Mellin transform of the SF. This can be 
considered as a benchmark solution, but it is unusable 
from the practical point of view for implementation 
in computational tools; 
\item exact MC solution, obtained by means of the PS 
approach, which is particularly powerful for 
exclusive event generation~\cite{psqcd,psqed,psohl}. The PS
method, originally developed and widely applied in 
perturbative QCD~\cite{psqcd}, has been 
recently introduced in QED~\cite{psqed,psohl} as a 
convenient framework to compute photonic radiative corrections 
in $e^+ e^-$ collisions. In fact, exponentiation of 
soft photons and the contribution of multiple emission of
hard collinear photons can be automatically accounted for.
\een
Let us summarize the basics of the PS method. 
The starting point of the PS approach is 
the Sudakov form factor~\cite{sff}:
\begin{equation} 
\Pi (s_1,s_2) = \exp \left[-\frac{\alpha}{2 \pi} 
\int_{s_2}^{s_1} \frac{d s'}{s'} \int_0^{x_+} dz P(z)  \right]  ,   
\label{eq:sudakov} 
\end{equation}
which represents the probability that an electron 
evolves from virtuality $-s_2$ to virtuality $-s_1$ with no 
emission of photons of energy fraction 
greater than $\epsilon = 1-x_+$, where 
$\epsilon$ is an infrared regulator. In terms of 
the factor $\Pi (s_1,s_2)$, 
the DGLAP equation can be written in 
iterative form as: 
\begin{eqnarray} 
D(x,s)&=&\Pi(s,m^2)\delta(1-x) \nonumber\\
&+& \int_{m^2}^s\Pi (s,s')\frac{d s'}{s'}\Pi (s',m^2)\frac{\alpha}{2\pi}
\int_0^{x_+} dy P(y) \delta (x-y) \nonumber\\
&+& \int_{m^2}^s\Pi (s,s')
\frac{ds'}{s'}\int_{m^2}^{s'}\Pi (s',s'')\frac{ds''}{s''}
\Pi (s'',m^2)\times \nonumber \\
&&\bigg(\frac{\alpha}{2\pi}\bigg)^2\int_0^{x_+} dx_1\int_0^{x_+}
dx_2 P(x_1)P(x_2) \delta (x-x_1x_2) + \cdots    
\label{eq:alpha2}             
\end{eqnarray}
where here $Q^2=s$ is understood, $s = 4 E^2$ being 
the total c.m. energy. 
Equation~(\ref{eq:alpha2}) suggests the steps to compute $D(x,s)$
by means of a MC algorithm~\cite{psqcd,psqed}:
\begin{figure}\bce
\epsfig{file=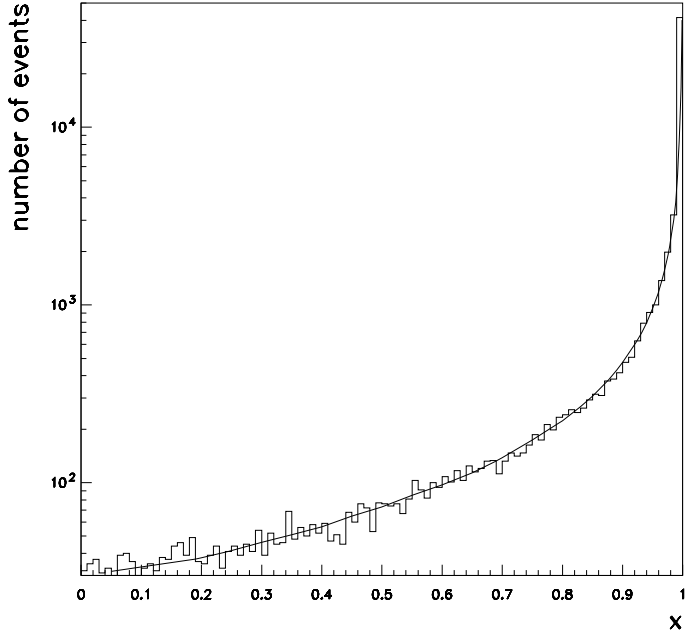,width=10cm} 
        \caption{$x$ distribution of the electron SF
        at $\sqrt{s} = 190$ GeV. Solid line: 
        numerical solution of DGLAP equation, by means of numerical 
        inversion of Mellin transform. Histogram: 
        result of the PS algorithm.}%
	\label{fig:1}
\ece\end{figure}	

\ben 
\item Set initial values for the electron virtuality and momentum fraction:
      $ K^2=m_e^2 $ and $x=1$ (with $m_e$ electron mass).  
\item Choose a random number $\xi$ in $[0,1]$ .
\item If $\xi<\Pi(s,K^2)$ then stop: no photon is radiated.
\item Else if $\xi>\Pi(s,K^2)$, a photon is emitted:
      calculate the value $ K^{\prime 2}$ of the electron 
      virtuality  after the branching as the solution of the equation     
      $\xi=\Pi(K^{\prime 2}, K^2)$. 
\item Choose the residual momentum fraction $z$ of the electron
      in $[0,x_+]$, according to $ P(z) $.  
\item Replace $x$ by $zx$ and  $K^2$ by $K^{\prime 2}$.   
      Go back to step 2.   
\een

\begin{figure}\bce
\epsfig{file=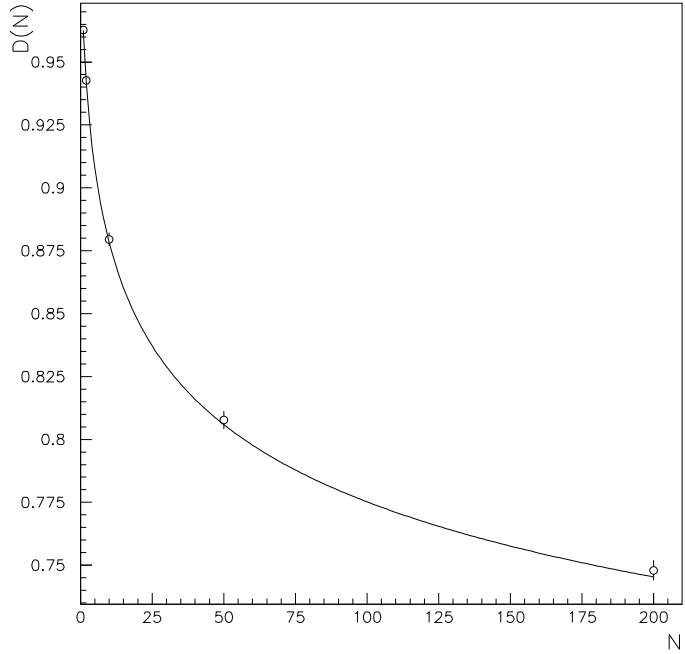,width=10cm} 
\caption{Comparison for the Mellin moments 
of the electron SF at $\sqrt{s} = 190$ GeV. 
Solid line: exact analytical moments. 
Markers: results of the PS algorithm for $N=1,2,10,50,200$.}
\label{fig:2}
\ece\end{figure}

\begin{figure}\bce
\epsfig{file=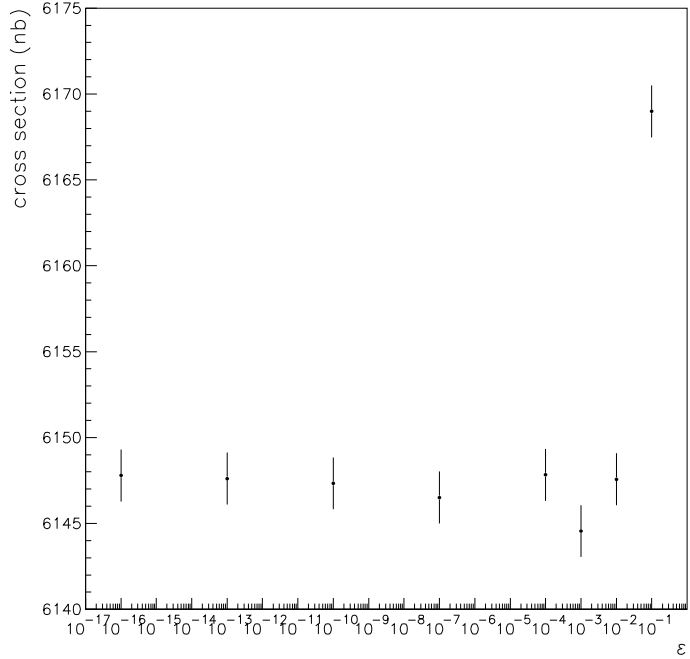,width=10cm} 
        \caption{QED corrected Bhabha cross section as a function of 
         the infrared regulator $\epsilon$, at the peak of the 
         $\Phi$ resonance. Cuts used are given in the text and correspond to 
         a realistic event selection at DA$\Phi$NE.}
\label{fig:3}
\ece\end{figure}

In this way the emission of  
a shower of photons by an electron is simulated, and 
the $x$ distribution of the PS event sample reproduces 
$D(x,Q^2)$. Such a distribution, 
obtained from a $10^5$ event sample 
at $Q^2 = s = (190)^2$ GeV$^2$, is compared in 
Fig.~\ref{fig:1} with a numerical solution of eq.~(\ref{eq:ap}),
normalized to the same number of events. The agreement 
is excellent. A further 
test of the algorithm is 
the comparison between the exact analytical Mellin moments 
$D(N)$ of the SF 
and the corresponding ones calculated in the PS scheme. This 
comparison is shown in Fig.~\ref{fig:2}, where it can be 
seen that the PS simulation (markers) for 
$N=1,2,10,50,200$ well agrees, within 
the statistical errors, with the analytical moments 
(solid line). 
The results shown in Fig.~\ref{fig:1} and Fig.~\ref{fig:2}
have been obtained 
with the value $\epsilon = 10^{-9}$ for the infrared 
cut-off entering eq.~(\ref{eq:sudakov}). 
However, independence of the PS predictions
for the QED corrected cross section 
from $\epsilon$ variation, on a wide range of
$\epsilon$ values (from $10^{-16}$ to $10^{-4}$), 
has been successfully checked, as shown in
Fig.~\ref{fig:3} for the LABH 
cross section at the $\Phi$ factory. In the present implementation 
of the PS model
$\epsilon$ is taken constant, in order to avoid loss of 
accuracy in the determination of the absolute value of the 
electron SF, as first pointed out in ref.~\cite{psqed}.

An up to $O(\alpha)$ PS algorithm has been developed as well. It allows
to calculate the corrected cross section of eq.~(\ref{eq:sezfs}) up to  
$O(\alpha)$. Such a calculation is strongly required for fully consistent 
comparisons between the PS predictions and an exact 
perturbative calculation. 
The steps required for 
the $O(\alpha)$ PS can be obtained by using 
eq.~(\ref{eq:alpha2}) and 
expanding  the product 
$D(x_-,Q^2)D(x_+,Q^2)D(y_+,Q^2)D(y_-,Q^2)$ present in 
eq.~(\ref{eq:sezfs}) up to $O(\alpha)$. It
is easy to see that:
\begin{eqnarray}
&&[D(x_-,s)D(x_+,s)D(y_+,s)D(y_-,s)]_{O(\alpha)}= \nonumber \\
&&[\Pi^4(s,m^2)]_{O(\alpha)} 
\delta(1-x_+)\delta(1-x_-)\delta(1-y_+)\delta(1-y_-)\nonumber \\
&&+\frac{\alpha}{2\pi}\ln\frac{s}{m^2}
\big\{\delta(1-x_-)\delta(1-y_+)\delta(1-y_-)P(x_+)\nonumber \\
&&\qquad \qquad \, \; +\delta(1-x_+)\delta(1-y_+)
\delta(1-y_-)P(x_-)\nonumber \\
&&\qquad \qquad \, \; +\delta(1-x_+)\delta(1-x_-)
\delta(1-y_+)P(y_-) \nonumber \\
&&\qquad \qquad \, \; +\delta(1-x_+)\delta(1-x_-)\delta(1-y_-)P(y_+)\big\},
\label{eq:4fsoal}
\end{eqnarray}   
where $[\Pi^4(s,m^2)]_{O(\alpha)}$ is the product of four Sudakov form
factors expanded up to $O(\alpha)$.

The steps for $O(\alpha)$ PS are therefore the following:
\ben 
\item Set initial values for the fermions virtuality and momentum fractions:
      $K^2=m_e^2$ and $x_+=x_-=y_+=y_-=1$.  
\item Choose a random number $\xi$ in $[0,1]$.
\item If $\xi<[\Pi^4(s,K^2)]_{O(\alpha)}$ then stop: no photon is radiated.
\item Else if $\xi>[\Pi^4(s,K^2)]_{O(\alpha)}$, a photon is emitted:
      calculate the value $ K^{\prime 2}$ of the fermion 
      virtuality  after the branching as the solution of the equation     
      $\xi=[\Pi^4(K^{\prime 2}, m^2)]_{O(\alpha)}$. 
\item Choose randomly the fermion which has emitted the photon. 
\item Choose the residual momentum fraction $z$ of the fermion
      in $[0,1-\epsilon]$, according to $ P(z) $ and replace $x_+,x_-,y_+$ 
      or
      $y_-$ with $z$, according to the particle which has radiated.  
\item Stop the algorithm.  
\een

The PS algorithm, both in the all order and $O(\alpha)$
implementation, offers the possibility to go naturally 
beyond the strictly collinear treatment of the electron evolution, 
by generating the transverse momentum $p_\perp$ of electrons and 
photons at each branching. To this end, a definite 
role for the variable $x$ must be chosen. 
In the QED PS model of ref.~\cite{psqed} $x$ is understood 
as the longitudinal momentum fraction of 
the electron after photon emission. In the present PS 
simulation, $x$ is chosen as 
the energy fraction of the electron after photon 
emission (hereafter denoted as $E$ scheme), in 
agreement with the known perturbative results for the 
photon spectrum in QED~\cite{ape} and previous interpretation  
in perturbative QCD (as, for example, in the paper by Marchesini-Webber 
in ref.~\cite{psqcd}). It is worth noticing that the two prescriptions are 
coincident in the collinear limit. In the $E$ scheme, the kinematics 
of the branching process $e(p) \to e'(p') + \gamma(q)$ can be written as:     
\begin{eqnarray}  
p&=&(E, \vec{0}, p_z) \nonumber\\ 
p'&=&(zE,  \vec{p}_\perp,  p'_z) \nonumber \\ 
q&=&((1-z)E,  - \vec{p}_\perp,  q_z) \, , 
\label{eq:kinealt} 
\end{eqnarray}  
with built-in energy and transverse momentum conservation. 
After having generated the variables
$k^2$,  ${k'}^2$ and $z$ by the PS algorithm, the on-shell
conditions $p^2=k^2$, ${p'}^2={k'}^2$, $q^2=0$, 
together with the longitudinal momentum con\-ser\-va\-tion, 
al\-low to obtain complete event reconstruction as follows: 
\eqnarray
&&p_z=E-\frac{k^2}{2E} \nonumber \\
&&p'_z=zE-\frac{(1-z)k^2+{k'}^2}{2E} \nonumber \\
&&q_z=(1-z)E-\frac{zk^2-{k'}^2}{2E} \nonumber \\
&&p^{2}_{\perp}=(1-z)(zk^2-{k'}^2) \, , \ \ 
\label{eq:kinep}
\endeqnarray
at first order in $k^2 / E^2 \ll 1$, $p^2_\perp / E^2 \ll 1$.
An alternative procedure, followed in the literature~\cite{psohl}, 
consists in generating a photon $p_{\perp}$ according to 
the leading pole behaviour $1/p \cdot q$. 
As a cross-check of the results obtained by means of eq.~(\ref{eq:kinep}), 
the method of ref.~\cite{psohl} has 
been also employed in the present implementation 
of the PS model, finding agreement between the procedures 
for the calculation of the QED corrected Bhabha cross section.

\begin{figure}\bce
\epsfig{file=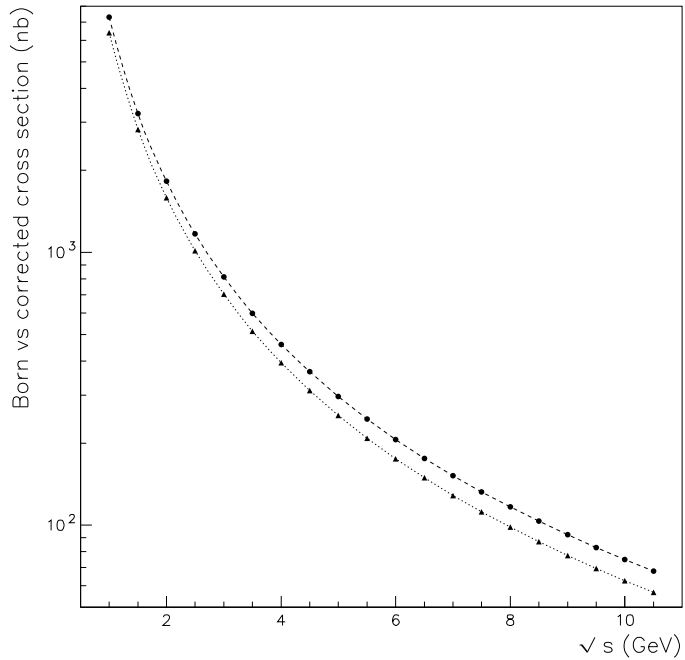,width=10cm} 
        \caption{Comparison between the QED corrected LABH cross section and 
        the Born one, in the c.m. energy range from 
	1 GeV to 10.5 GeV. Solid circles: Born approximation; solid triangles:
	QED corrected cross section. Cuts are given in the text.}%
	\label{fig:4}
\ece\end{figure}	
\begin{figure}\bce
\epsfig{file=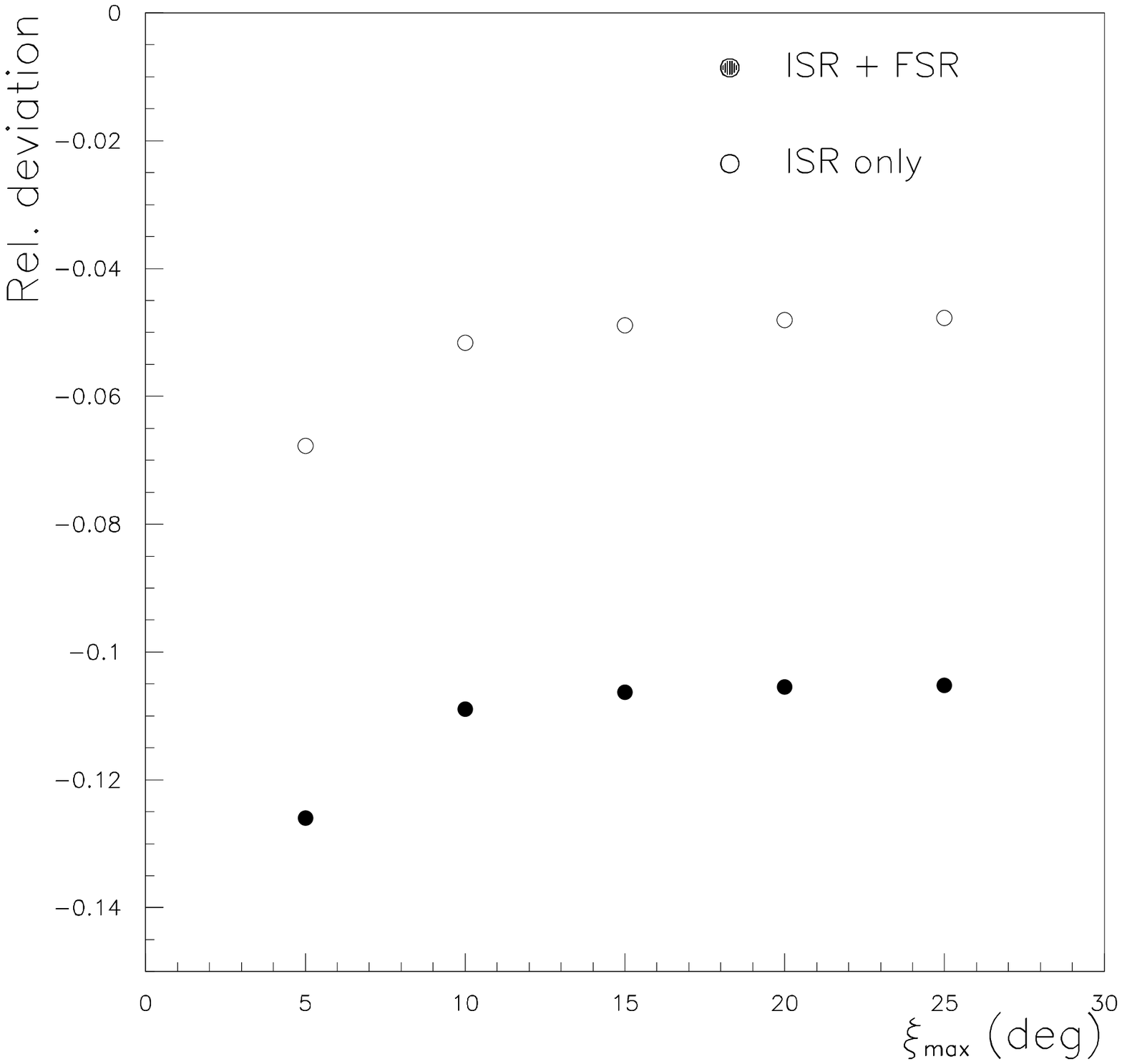,width=10cm} 
        \caption{Relative effect of ISR and 
        ISR+FSR on the integrated Bhabha cross section at the 
        $\Phi$ factories. 
        Open circles: ISR only; solid circles: 
        ISR+FSR. Cuts used are given in the text.}%
	\label{fig:5}
\ece\end{figure}	

\section{Bhabha generator and first sample of phenomenological 
results}
In the spirit of the PS approach described above, a new MC 
event generator (BABAYAGA) for simulation of the LABH
process at $e^+ e^-$ flavour factories has been 
developed. It is a generator of unweighted events, 
giving as output the QED corrected cross section 
and the momenta of the final-state particles. Both ISR and 
FSR are simulated. In principle, an ``arbitrary'' 
number of photons can be generated, including their 
$p_\perp$. In the standard version, BABAYAGA 
records the momenta of electron and positron, and of 
the most energetic and next-to-most energetic
photons generated by the electromagnetic shower. The possibility 
of an up to $O(\alpha)$ calculation 
of eq.~(\ref{eq:sezfs}) is included as an option, 
in order to compare it with the exact $O(\alpha)$ 
perturbative results (see Sect.~5). Also for 
the $O(\alpha)$ branch, particles 
momenta are reconstructed. 

As a first example, in Fig.~\ref{fig:4}, the PS corrected 
LABH cross section is
compared with the Born-like cross section, as a function of the c.m. energy. An 
energy threshold of $0.8\cdot E_{beam}$ for both electron and positron 
is required, the acceptance cuts are
$20^\circ\leq \vartheta_{\pm}\leq 160^\circ$ and an acollinearity cut of 
$5^\circ$ is imposed. In such a situation, the correction due to QED 
radiation
is of the order of 10\% at the $\Phi$ factories and of the order 
of 20\% at the
$B$ factories, pointing out the need of a careful treatment of photonic
corrections in simulation tools of the LABH  process at flavour factories. 
\begin{figure}\bce
\epsfig{file=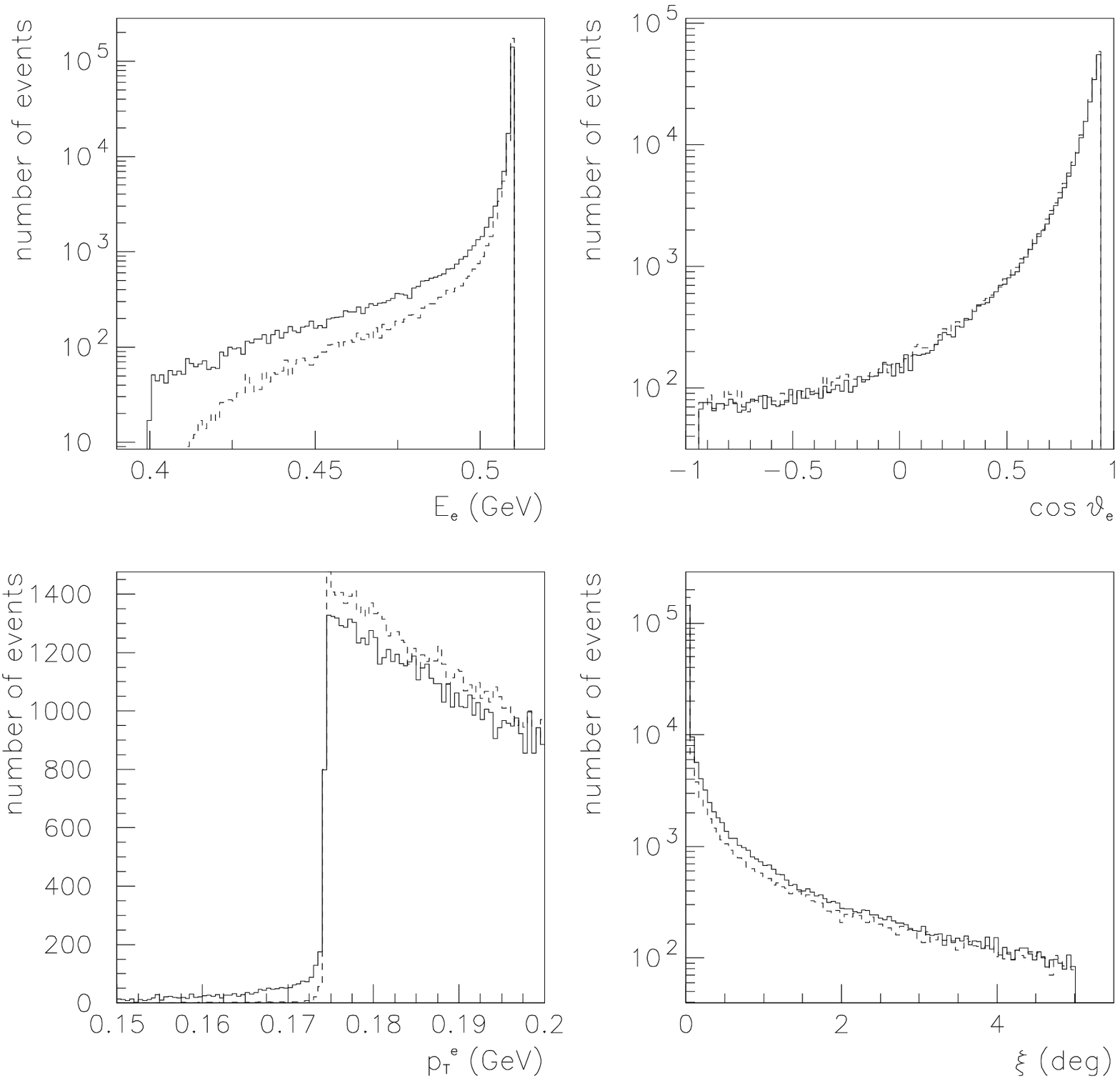,width=10cm} 
        \caption{Effect of FSR on Bhabha differential distributions 
        at the $\Phi$ factories:
        electron energy, electron scattering angle, electron transverse 
        momentum and acollinearity.
        Dashed histograms: ISR only. 
        Solid histograms: ISR+FSR. Cuts are given in the text.}%
	\label{fig:6}
\ece\end{figure}	

\begin{figure}\bce
\epsfig{file=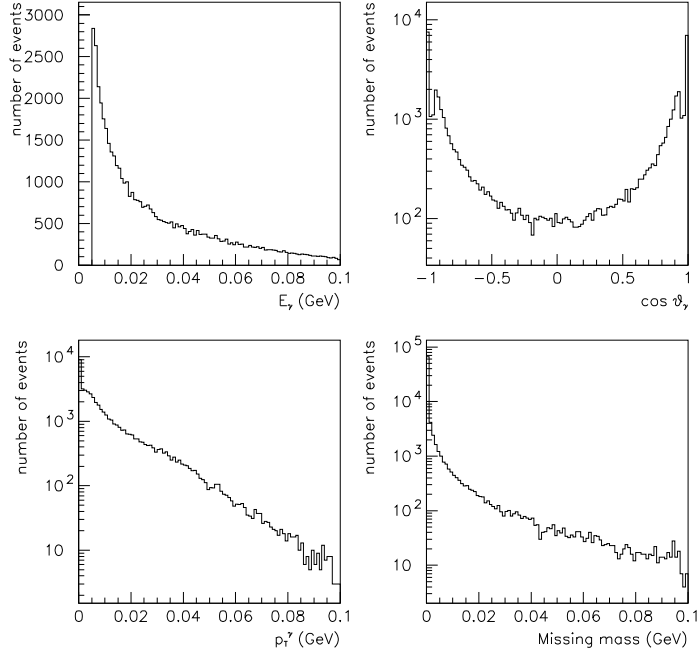,width=10cm} 
\caption{Distributions of the most-energetic photon 
in the LABH process at the $\Phi$ factories: energy, polar angle, 
transverse momentum and missing mass. Cuts are given in the text.}%
\label{fig:7}
\ece\end{figure}

A further sample of phenomenological results obtained by means of 
the BABAYAGA program is shown in Figs.~\ref{fig:5}-\ref{fig:7}. 
The parameters and selection criteria adopted in the 
analysis are very similar to those 
considered in previous MC simulations~\cite{cv,dv} and correspond 
to realistic data taking at DA$\Phi$NE and 
VEPP-2M, for c.m. energy at the $\Phi$ peak, 
i.e. $\sqrt{s} = 1.019$~GeV. The energy cut imposed on the 
final-state electron and positron is $E_{min}^{\pm} = 0.4$~GeV; 
two different angular acceptances 
of $20^\circ \leq \vartheta_{\pm} \leq 160^\circ$ and 
$50^\circ \leq \vartheta_{\pm} \leq 130^\circ$ are 
considered, with (maximum) 
acollinearity cut allowed to vary in the range
$\xi_{max} = 5^\circ, 10^\circ, 15^\circ, 
20^\circ,25^\circ$. The 
energy cut refers to the energy of the bare 
electron and positron, corresponding to 
a so-called bare ES~(see 
for instance ref.~\cite{lep2bha}), which is not far from 
realistic due to the presence of magnetic fields as for
DA$\Phi$NE and CMD-2 experiment at VEPP-2M. The tight 
acollinearity cut $\xi_{max} = 5^\circ$ is generally introduced 
in MC simulations in order to single out quasi-elastic Bhabha 
events~\cite{cv}. However, it is worth 
noticing that this acollinearity cut, in association with 
the high energy thresholds on bare electrons and 
positrons, defines a rather severe set of constraints, 
which can be expected to emphasize the effects of 
QED radiative corrections, especially from the 
final-state. This conjecture is confirmed by the numerical 
results of Fig.~\ref{fig:5}, where the (relative) effect, with respect to 
the Born cross section, of 
ISR only (open circles) is compared with the whole 
effect of ISR and FSR (solid circles). Actually, it can be 
seen that the photon radiation produces a 
lowering of the integrated cross section of the 
order of 10\% and that half of this effect has to be ascribed 
to FSR, showing the need of including FSR for realistic
simulations. In this simulation, 
the $Q^2$ entering the electron SF is fixed to be $Q^2=s$ as 
typical virtuality 
in initial- and final-state shower. The relevance of FSR 
in such ES is further illustrated in Fig.~\ref{fig:6} 
for the electron energy, electron scattering angle, 
electron transverse momentum distribution and acollinearity 
distribution as well. It can be noticed that the electron 
energy and $p_{\perp}$ are significantly affected by FSR, 
while this is not the case for the electron scattering angle 
and acollinearity, which are actually 
largely dominated, as expected, by the $x$ distribution of 
ISR~\cite{prd}.
In the comparison shown in Fig.~\ref{fig:6} 
the numbers of generated events for ISR only and ISR+FSR 
are consistently normalized to the same luminosity.

A further illustration of the potential of BABAYAGA in 
event generation is given in Fig.~\ref{fig:7}, showing  
the energy, polar angle and transverse momentum 
of the most energetic photon, 
as well as the missing mass distribution of the event, defined as 
$\sqrt{(p_+ + p_- - q_+ - q_- -k)^2}$, where $p_{\pm}(q_{\pm})$ are 
the initial(final) electron/positron momenta and 
$k$ is the momentum of the most energetic photon. 
A minimum energy cut of 5~MeV is imposed as visibility criterion. 
It can be seen that the expected physical features of photon 
radiation are correctly reproduced by the PS generator, 
noticeably the soft peaking behaviour of $E_\gamma,
p_{\perp}^{\gamma}$ and missing 
mass distribution as well as the initial- and 
the final-state collinear peaks, which are clearly visible 
in the $\cos\vartheta_\gamma$ distribution.   
	
\section{Benchmark calculation}
In order to test the (physical+technical) precision of the PS approach 
and corresponding Bhabha generator, an {\em exact} $O(\alpha)$ 
perturbative calculation has been addressed, by computing the 
up to $O(\alpha)$ corrected cross section as follows
\begin{equation}
\sigma_{exact}^{(\alpha)} = \sigma_{S+V}^{(\alpha)} (E_\gamma < k_0) 
+ \sigma_{H}^{(\alpha)} (E_\gamma > k_0, cuts) .
\label{eq:exact}
\end{equation}
In the above equation the $O(\alpha)$ soft+virtual part 
$\sigma_{S+V}^{(\alpha)}$ is obtained by integration over the 
electron angle of the soft+virtual differential 
cross section, which is explicitly given by~\cite{vs}
\begin{equation}
d \sigma_{S+V}^{(\alpha), i} = d \sigma_0 
\left\{ 1 + 2\, \left( \beta_e + \beta_{int} \right) 
\ln k_0/E + C_F^i \right\} .
\label{eq:sv}
\end{equation}
In eqs.~(\ref{eq:exact}) and (\ref{eq:sv}), $k_0$ stands for a 
(small) photon energy soft-hard (fictitious) separator, 
$i$ is an index for the 
photonic contributions to the Bhabha Born matrix element ($i = 
|\gamma(s)|^2$,$|\gamma(t)|^2$,$\gamma(s)$-$\gamma(t)$),
$\beta_e = 2 \alpha / \pi 
\left(\ln(s/m_e^2) - 1 \right)$ is the leading 
collinear factor for initial- and final-state radiation, 
$\beta_{int} = 2 \alpha / \pi \ln(t/u)$ is
the leading angular factor for initial-final state interference. 
Furthermore, in eq.~(\ref{eq:sv}) 
$C_F^i = C_F^i(\vartheta)$ are the 
soft+virtual $K$-factors for the three QED channels, 
including box/interference finite terms~\cite{vs}. 
The hard bremsstrahlung contribution $\sigma_{H}^{(\alpha)}$ 
is included, in the photon phase-space region 
above the energy cut $k_0$, via 
the $e^+ e^- \to e^+ e^- \gamma$ matrix element
calculated in ref.~\cite{hard}. Its contribution is computed 
numerically in the presence of experimental cuts
by means of the MC method with standard 
importance sampling technique,  
in order to handle collinear and infrared 
photon singularities. Needless to say, the independence 
of eq.~(\ref{eq:exact}) 
from the fictitious infrared cut-off $k_0$ 
was successfully checked with high numerical precision.

The comparison between the exact $O(\alpha)$ 
calculation and the $O(\alpha)$ predictions 
of the PS generator is of interest because it allows to 
evaluate the size of the $O(\alpha)$ next-to-leading order
(NLO) corrections, which are missing in the PS.
Moreover, this comparison can be a useful guideline 
to improve the agreement between perturbative and LL PS results, 
for example, by properly choosing the virtuality $Q^2$ 
in the electron SF in such 
a way that $O(\alpha)$ NLO terms are 
effectively reabsorbed into the 
LL contributions. It can be noticed, in fact, 
that by choosing the scale $Q^2$ as $Q^2 = st/u$ in the collinear 
logs $\ln(Q^2/m^2)$ generated by the PS method, then 
one has that $\ln(Q^2/m^2) \to 
\ln(s/m^2) + \ln(t/u)$~\cite{stu}. If this 
procedure is applied both to ISR and FSR by choosing 
as maximum virtuality of the electromagnetic shower 
$Q^2 = st/u$, it is possible to keep under control, 
besides the large logarithms from ISR and FSR, also the leading 
angular contribution from initial-final state interference.

In order to estimate the precision of the PS approach 
with all order corrections, higher-order LL terms must  
be added to the exact $O(\alpha)$ cross section 
in the benchmark calculation. The general algorithm 
recently proposed in ref.~\cite{a2l}, and 
there applied to the high-precision 
computation of the SABH cross section, 
can be advocated, by writing the benchmark cross section as 
(in the so-called additive form)~\cite{a2l}
\begin{equation}
\sigma = \sigma_{LL}^{(\infty)} - \sigma^{(\alpha)}_{LL} 
+ \sigma^{(\alpha)}_{exact} ,
\label{eq:add}
\end{equation}
where $\sigma_{LL}^{(\infty)}$ is the 
all-order LL cross section as given by eq.~(\ref{eq:sezfs}),  
$\sigma^{(\alpha)}_{LL}$ is the up to $O(\alpha)$ 
truncation of the 
LL cross section, $\sigma^{(\alpha)}_{exact}$ 
is the exact perturbative 
$O(\alpha)$ cross section of eq.~(\ref{eq:exact}).
Collinear SFs as given in ref.~\cite{sfcoll} are 
used in the calculation of the LL cross sections.
Adopting such a procedure,
 exact $O(\alpha)$  corrections are 
simply matched with LL higher-orders in the collinear approximation. 
Therefore, the cross section 
of eq.~(\ref{eq:add}) includes exact $O(\alpha)$ + $O 
(\alpha^n L^n$), with $n \geq 2$, leading corrections.

A more accurate factorized form, as 
motivated and 
discussed in detail in ref.~\cite{a2l}, can be supplied, 
by computing the cross section as
\begin{eqnarray}
&&\sigma_F  \simeq ( 1 + C_{NL}^{(\alpha)} ) \, 
\sigma_{LL} , \nonumber \\
&&\nonumber \\
&&C_{NL}^{(\alpha)} \equiv \frac{\sigma^{(\alpha)}_{exact}
- \sigma^{(\alpha)}_{LL} }{\sigma_0} \equiv
\frac{\sigma_{NL}^{(\alpha)}}{\sigma_0} ,
\label{eq:fact}
\end{eqnarray}
where $C_{NL}^{(\alpha)}$ is the whole NLO content of 
eq.~(\ref{eq:exact}). In such a way, the bulk of the most 
important second-order 
NLO corrections, i.e. the 
$O (\alpha^2 L)$ terms, are added to the content 
of eq.~(\ref{eq:add})~\cite{a2l}\footnote{Actually, in 
ref.~\cite{a2l} a more refined
treatment of the factorized cross section is given, and it is shown that
eq.~(\ref{eq:fact}) is an approximation of the full treatment at the 0.1\%
level, which is sufficient for the present study.}.

The procedure here shortly sketched 
works efficiently for cross section calculation 
(not unweighted event generation) and allows, 
by comparison with the full PS predictions, 
to estimate the overall precision of 
the PS approach. Furthermore, it is 
possible to get a measure of $O(\alpha^n L^n)$ and 
$O(\alpha^2 L)$ corrections. The benchmark calculation 
is available in the form of MC integrator program 
(LABSPV), which is a suitable modification of 
the SABSPV code~\cite{sabspv} from SABH to 
low-energy LABH process.

\section{Tests of the approach and further numerical results}
\begin{figure}\bce
\vspace{8.0cm}
\includegraphics{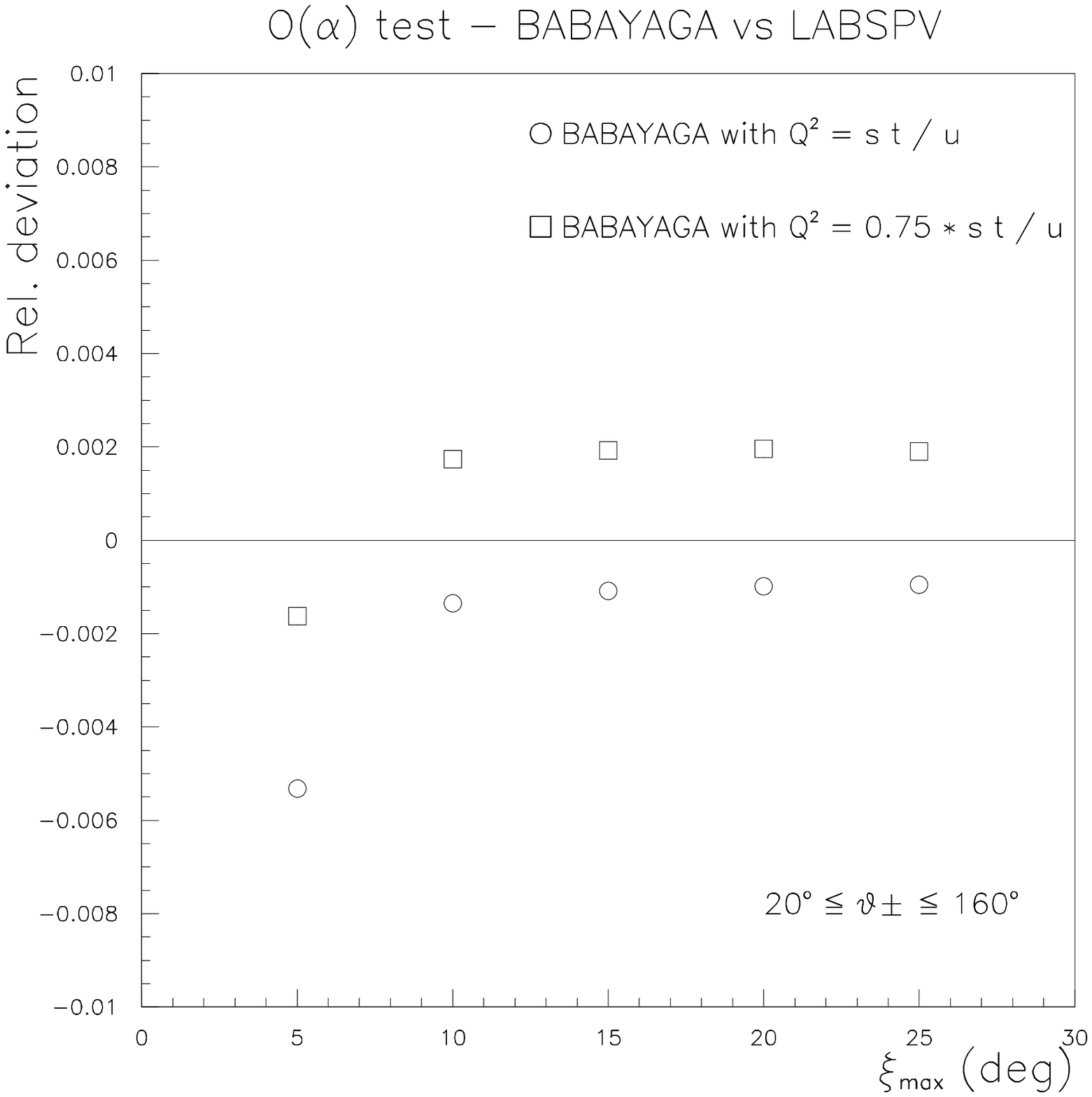}
        \includegraphics{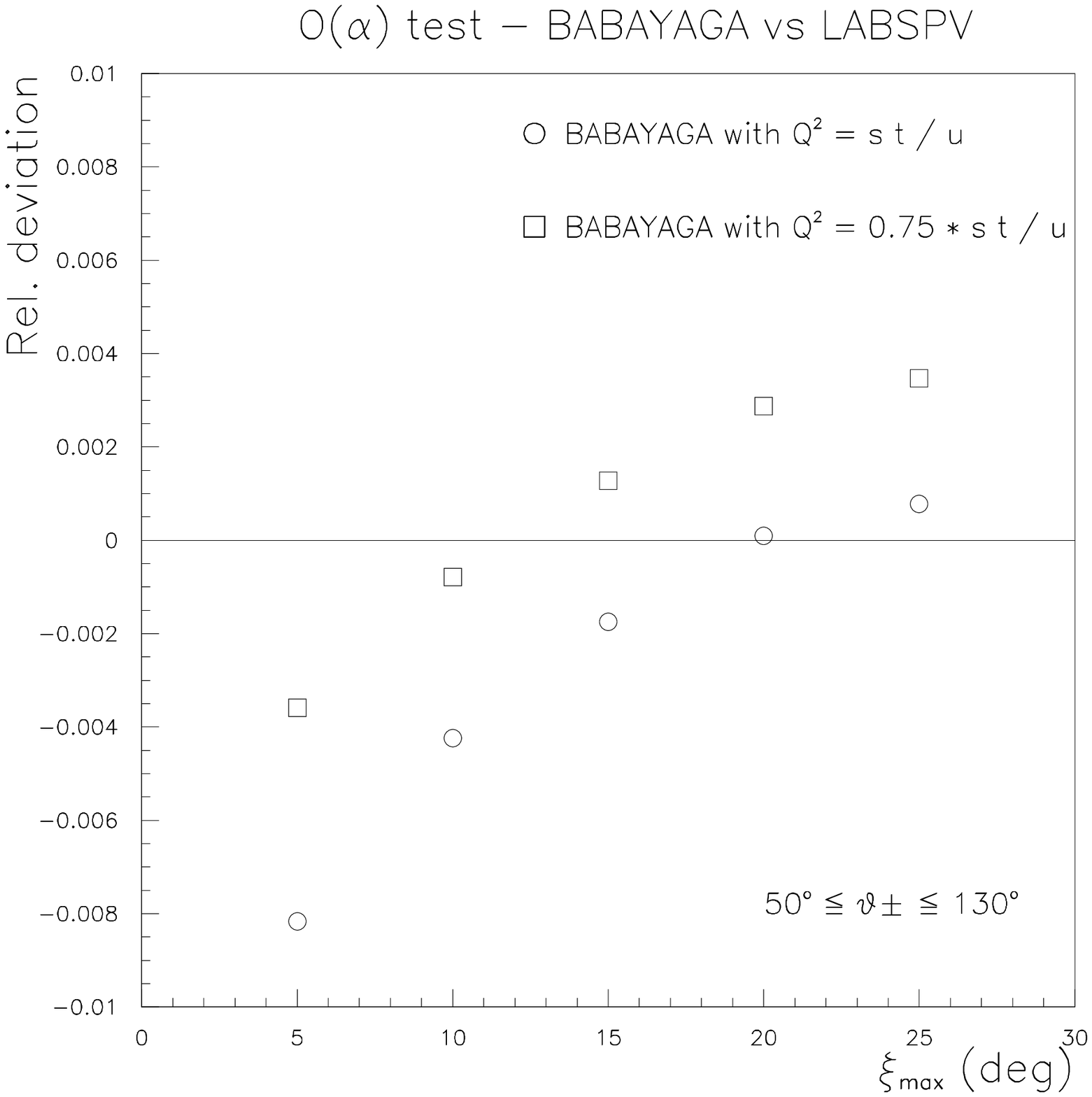}
	 \caption{Relative differences for the LABH process at the 
        $\Phi$ factories 
        between the exact $O(\alpha)$ cross section (LABSPV) and 
        the up to $O(\alpha)$ PS one (BABAYAGA), as functions 
        of the acollinearity cut and for two choices of the 
        $Q^2$ scale in the electron SF. On the left the
	acceptance region $20^\circ\leq\vartheta_{\pm}\leq 160^\circ$ 
        is considered, 
	while on the right the
	acceptance region is $50^\circ\leq\vartheta_{\pm}\leq  130^\circ$. 
        Other cuts are specified in the text.}%
	\label{fig:8:9}
\ece\end{figure}

As already remarked in the previous section, 
the detailed comparison between the 
exact $O(\alpha)$ cross section of 
eq.~(\ref{eq:exact}) and the up to $O(\alpha)$ 
expansion of the PS results is a valuable tool to establish the 
physical precision of the PS approach, since the 
registered difference is due to the NLO corrections left over 
in the pure LL predictions of the PS scheme. The relative difference
between eq.~(\ref{eq:exact}) and the up to $O(\alpha)$  
PS cross section is shown in Fig.~\ref{fig:8:9} for the 
angular acceptances $20^\circ \leq \vartheta_{\pm} 
\leq 160^\circ$ and $50^\circ \leq \vartheta_{\pm}\leq 130^\circ$ and for 
two different choices of the $Q^2$ scale in the PS, i.e. for
$Q^2 = s t/u$ and $Q^2 = 0.75 \cdot s t/u$. The relative deviations 
shown in the figure are plotted as functions of the acollinearity cut.  
For  
the ``natural'' choice of the scale $Q^2 = s t/u$ which allows to 
completely reproduce the LL structure 
of the exact $O(\alpha)$ calculation, 
the difference is an unambiguous evaluation of the 
NLO corrections. Such contributions, as a priori expected, are important 
in view of the required theoretical precision, especially at 
the acollinearity value $\xi_{max} =
5^\circ$, being of the order of 
several 0.1\% (see Fig. \ref{fig:8:9}). 
However, it is worth noticing that a simple 
variation of the $Q^2$ scale from $Q^2 = s t/u$ to 
$Q^2 = 0.75 \cdot s t/u$ significantly reduces the difference, which 
goes from more than the 0.5-1\% level down to about 0.1-0.3\% at $\xi_{max} 
= 5^\circ$, depending on the angular set up. In general, 
with the adjusted scale 
$Q^2 = 0.75 \cdot s t/u$ 
the difference between the exact $O(\alpha)$ calculation 
and the PS predictions is within 0.5\%. This naive example 
illustrates how, for a given set of cuts, the level of agreement   
can be substantially improved by a simple redefinition of 
the maximum virtuality of the 
electromagnetic shower. Going beyond this simple procedure would require a 
merging between perturbative calculation and 
PS scheme, which is beyond the scope of the present work.

\begin{figure}\bce
\vspace{8.0cm}
\includegraphics{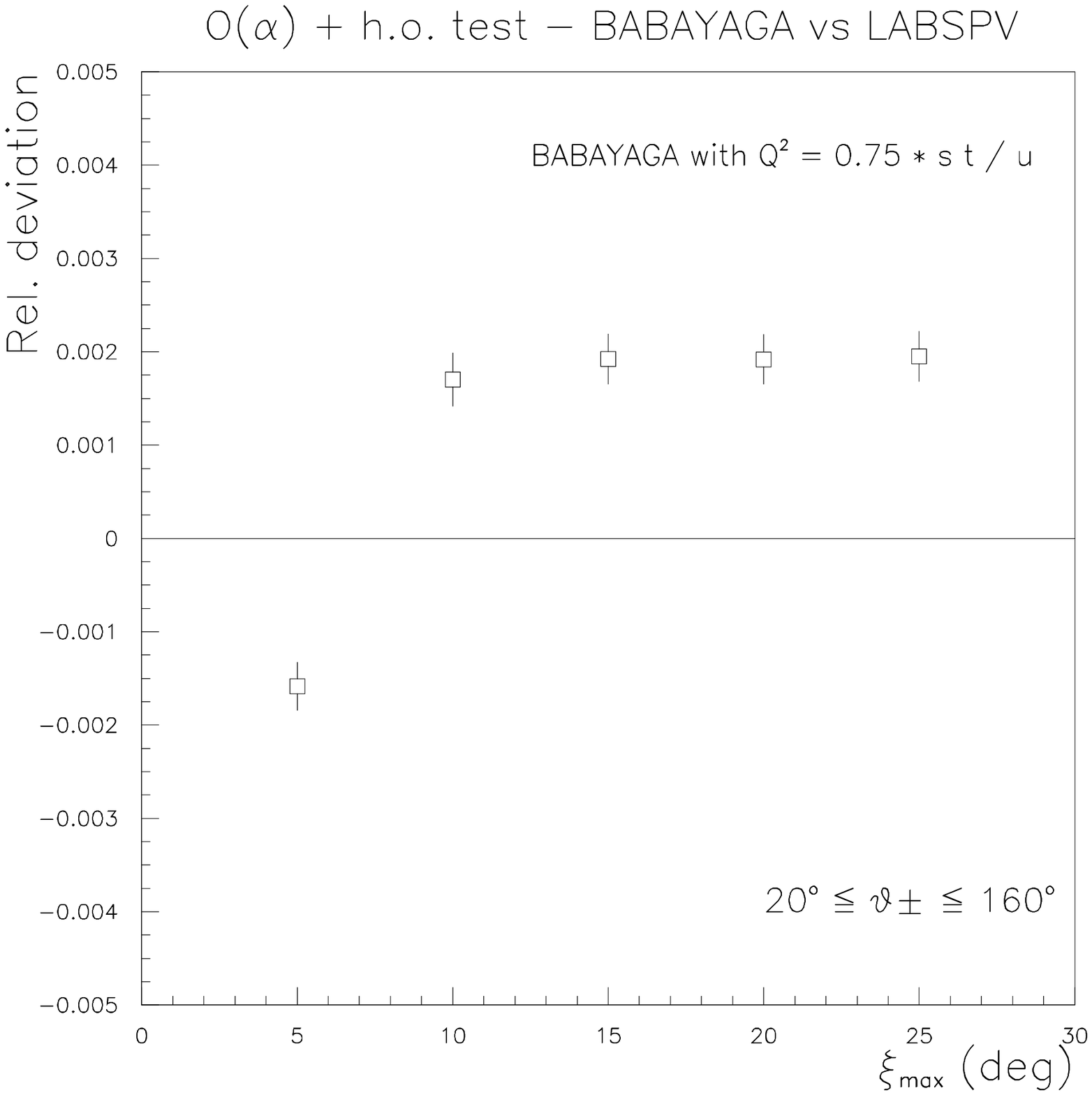}
        \includegraphics{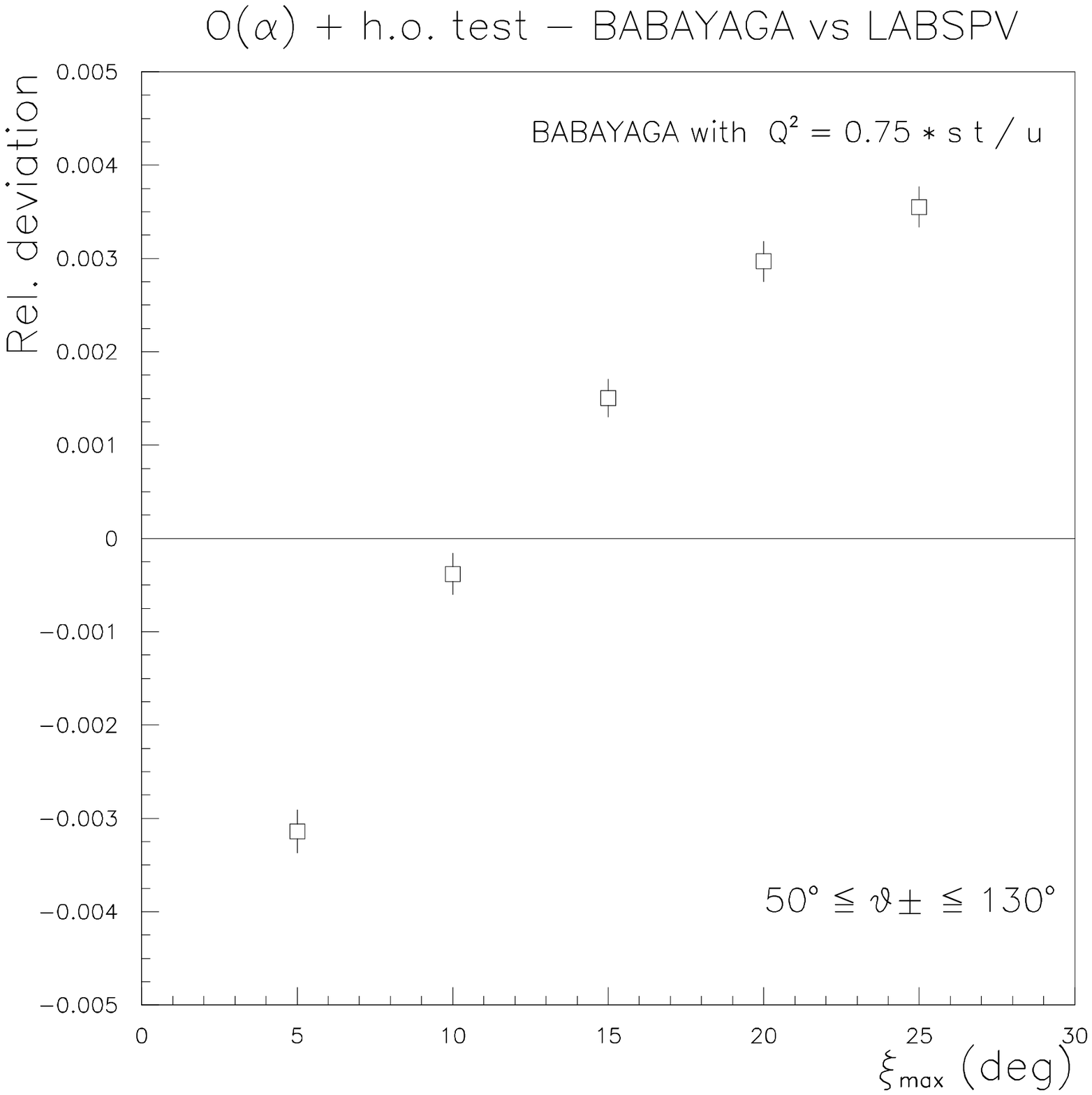}
	 \caption{The same as Fig.~\ref{fig:8:9} for 
        the $O(\alpha)$ plus higher orders Bhabha cross section at the 
        $\Phi$ factories. Cuts are given in the text.}%
	\label{fig:10:11}
\ece\end{figure}

In order to assess the reliability of the PS approach 
in the presence of higher-order corrections, the comparison 
between the benchmark and PS calculation has been extended 
to the computation of 
the exact $O(\alpha)$ plus higher-orders cross section.
The relative difference
between eq.~(\ref{eq:add}) and the full   
PS cross section is shown in Fig.~\ref{fig:10:11}, still
as functions of the acollinearity cut. According to what 
discussed about the $O(\alpha)$ comparison, 
the scale in the PS is, for definiteness, fixed at 
$Q^2 = 0.75 \cdot st/u$. Relative differences contained within 
0.5\% are still
observed, confirming the equivalent 
implementation of higher-order LL corrections in the PS and 
benchmark calculation. This difference between 
BABAYAGA and LABSPV in the presence of higher-order corrections 
can be considered as an estimate of the physical 
precision of the ``modified'' PS approach for the cross section 
calculation for realistic ES at the $\Phi$ factories. 

\begin{figure}\bce
\vspace{8.0cm}
\includegraphics{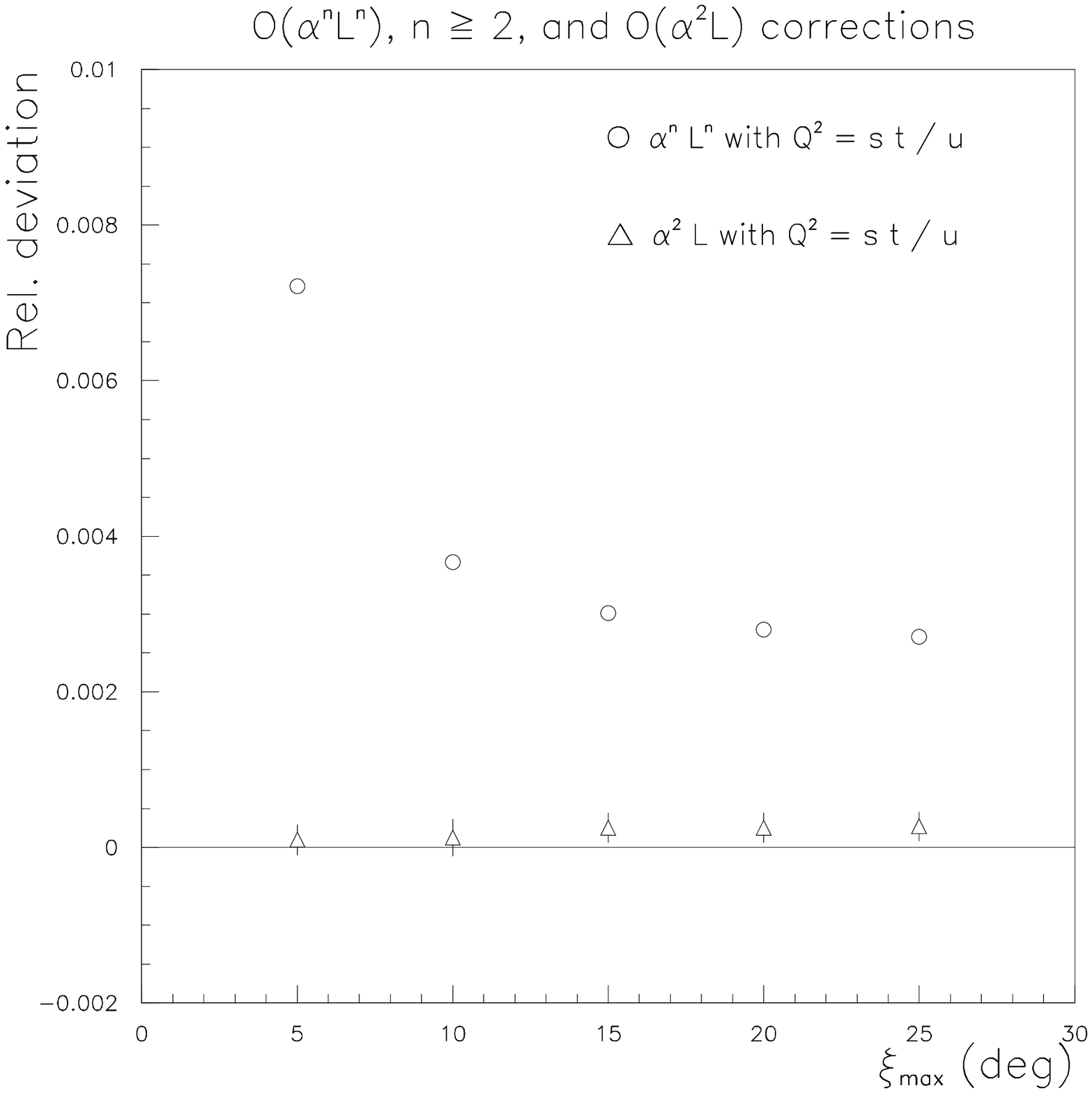}
        \includegraphics{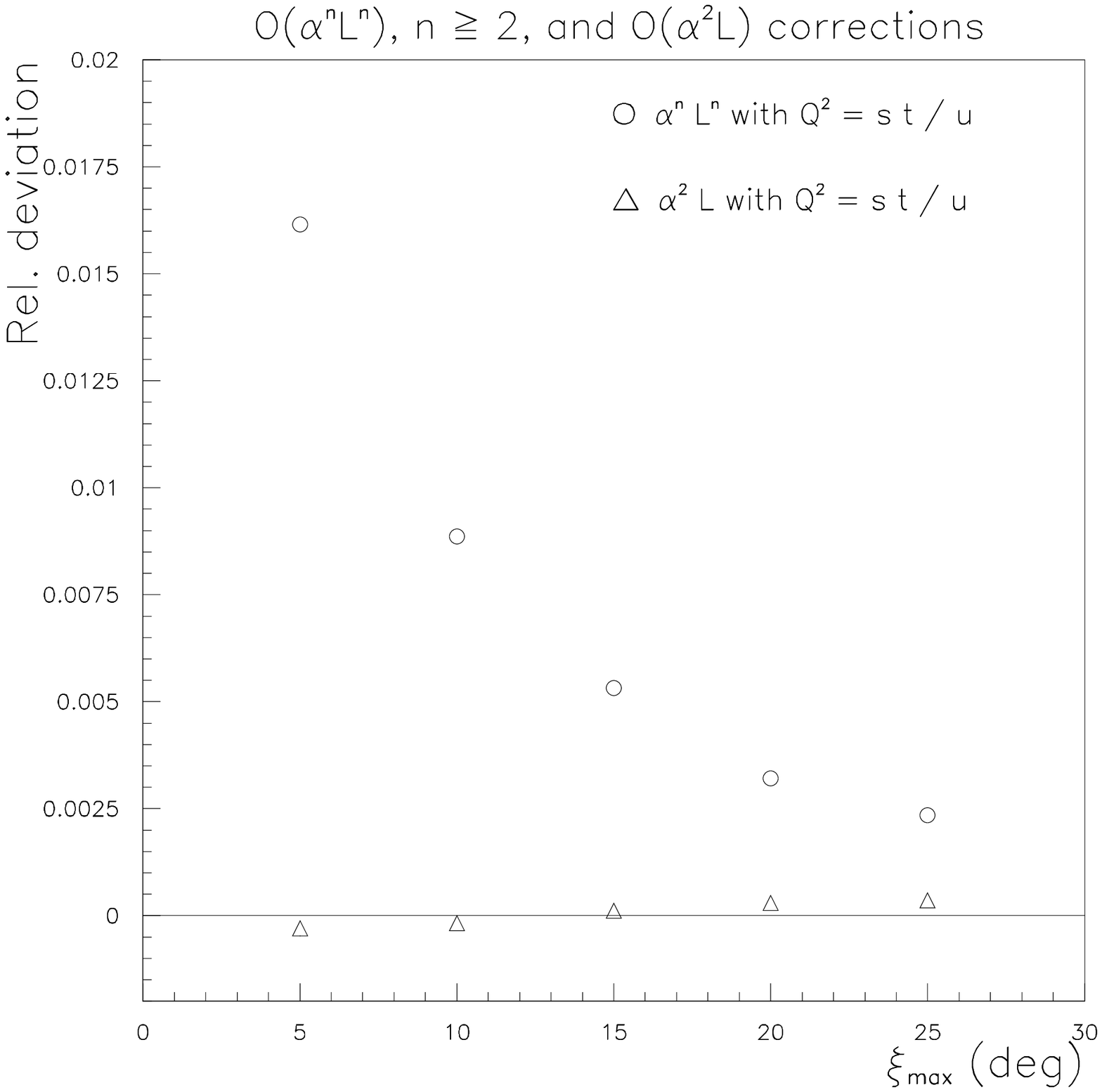}
	 \caption{Effect of higher-order
        $O (\alpha^n L^n$) ($n \geq 2$) and 
        $O (\alpha^2 L$) corrections on the 
        integrated Bhabha cross section at the $\Phi$ factories. On the left the
	acceptance region $20^\circ\leq\vartheta_{\pm}\leq 160^\circ$ 
	is considered, 
	while on the right the
	acceptance region is $50^\circ\leq\vartheta_{\pm}\leq  130^\circ$. 
        Other cuts are specified in the text.}%
	\label{fig:12:13}
\ece\end{figure}

In addition to the evaluation of the $O(\alpha)$ NLO 
corrections, for an assessment of the 
theoretical precision, it is important to evaluate the impact 
of the higher-order LL contributions. The 
size of LL $O (\alpha^n L^n$), with $n \geq 2$, 
corrections can be derived, as already 
remarked, by comparing in the benchmark calculation the results of 
eq.~(\ref{eq:exact}) and eq.~(\ref{eq:add}) 
or, equivalently, in the PS scheme the full all-order predictions 
with the corresponding up to $O(\alpha)$ truncation. Furthermore, 
the difference between eq.~(\ref{eq:add}) and eq.~(\ref{eq:fact})
in the benchmark calculation is able to provide an indication   
of the size of $O (\alpha^2 L)$ corrections, thus yielding 
an estimate of the most important second-order 
NLO corrections left over 
in the PS method. Such an analysis of higher-order effects 
leads to the results shown in Fig.~\ref{fig:12:13}, where 
the relative effect of the above higher-order corrections is 
shown for the natural scale $Q^2 = s t/u$. It can be clearly seen that 
LL $O (\alpha^n L^n$) corrections are unavoidable 
in view of the expected theoretical accuracy. In fact, 
their contribution is at 1-2\% level at $\xi_{max} = 
5^\circ$ and of the order of some 0.1\% for larger acollinearity cuts. The 
impact of $O (\alpha^2 L$) is, instead, negligible, being 
well below the 0.1\% level, in agreement with the 
estimate given in ref.~\cite{abaetal}.   

\begin{figure}\bce
\epsfig{file=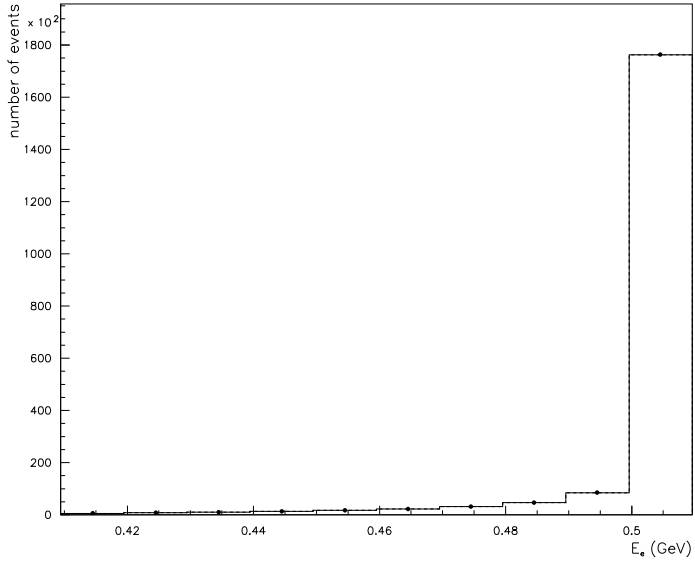,width=10cm} 
\caption{Electron energy distribution for the LABH 
process at the $\Phi$ factories. Markers: exact $O(\alpha)$ via LABSPV.
Histograms: PS prediction via BABAYAGA with scale $Q^2 = 0.75 \cdot st/u$}%
\label{fig:14}
\ece\end{figure}

\begin{figure}\bce
\epsfig{file=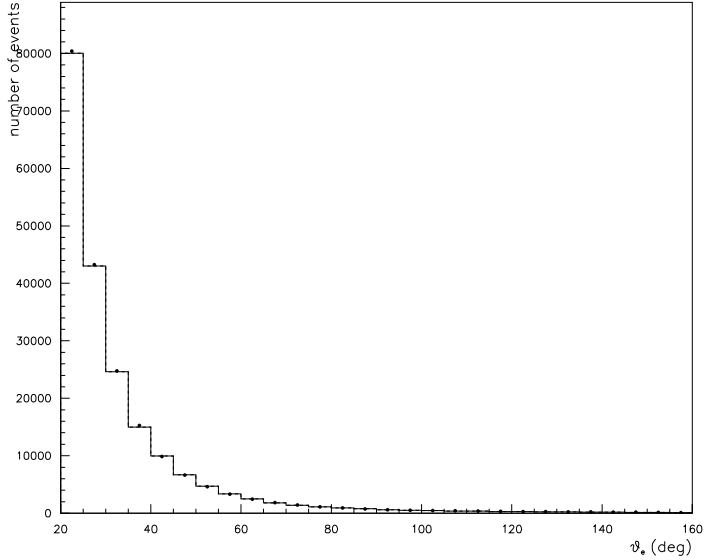,width=10cm} 
\caption{Electron angle distribution for the LABH 
process at the $\Phi$ factories. Markers: exact $O(\alpha)$ via LABSPV.
Histograms: PS prediction via BABAYAGA 
with scale $Q^2 = 0.75 \cdot st/u$}%
\label{fig:15}
\ece\end{figure}

\begin{figure}\bce
\epsfig{file=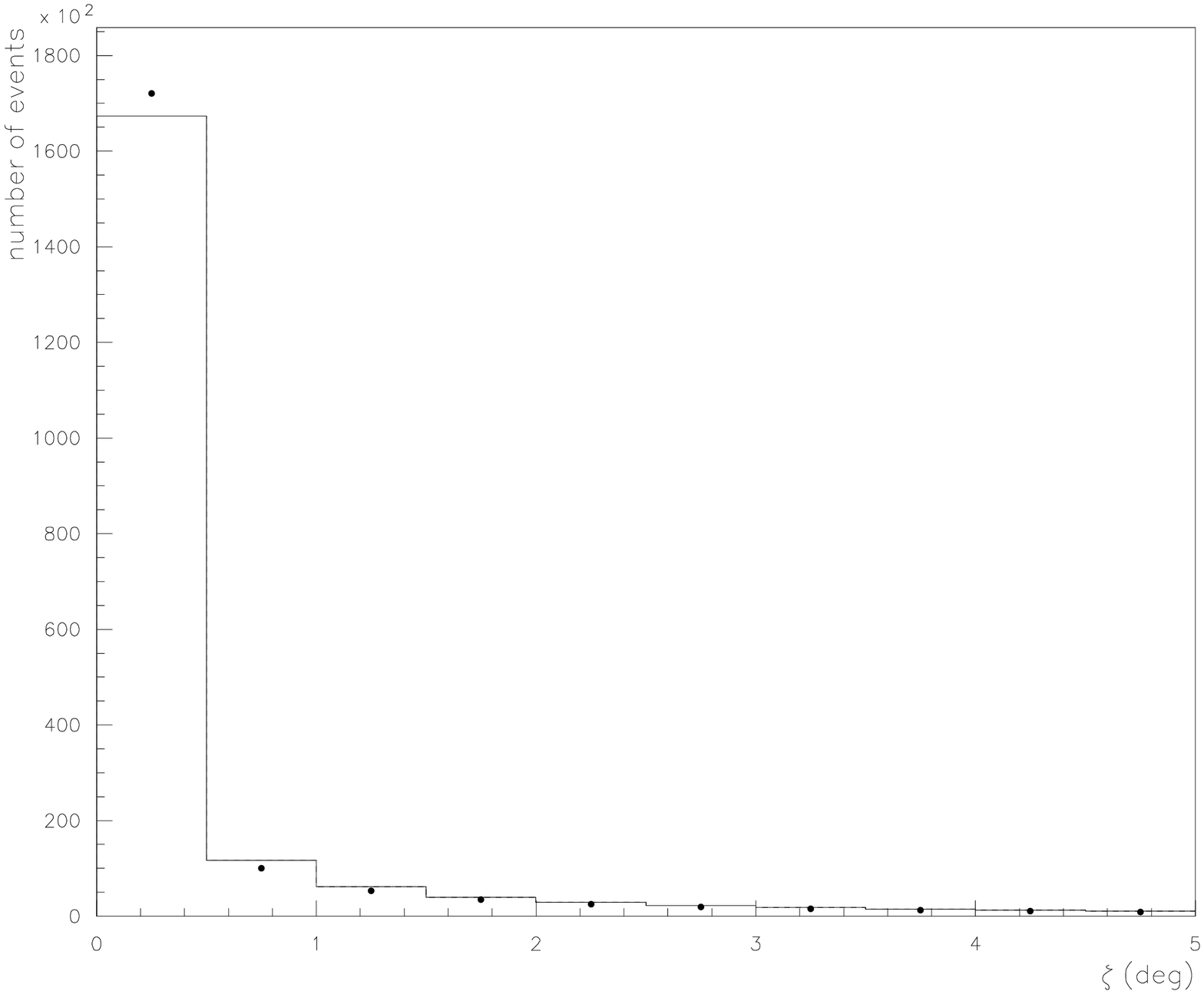,width=10cm} 
        \caption{Acollinearity angle distribution 
        for the LABH 
        process at the $\Phi$ factories.
        Markers: exact $O(\alpha)$ via LABSPV.
        Histograms: PS prediction via BABAYAGA 
        with scale $Q^2 = 0.75 \cdot st/u$}%
	\label{fig:16}
\ece\end{figure}

Having established that the inclusion of higher-order contributions
does not alter, as expected, the results of the 
$O(\alpha)$ comparison, further 
$O(\alpha)$ tests have been performed at the 
level of exclusive differential distributions. The 
results are illustrated 
in Figs.~\ref{fig:14}-\ref{fig:16}. In these plots, 
the histograms represent the distributions of 
the events generated by means 
of the up to $O(\alpha)$ PS generator BABAYAGA with scale 
$Q^2 = 0.75 \cdot st/u$, while the markers correspond to the 
predictions of the benchmark calculation LABSPV. A few  
comments are in order here. Since the benchmark calculation
is not suited for unweighted event generation, the number of events 
corresponding to the markers have been obtained by calculating, 
as a first step, the integrated cross section over each bin and 
next by multiplying it for a reference luminosity  
calculated as $L = N_{PS}/\sigma_{PS}$, where 
$N_{PS}$ and $\sigma_{PS}$ are the number of events and the 
cross section obtained with the PS generator, respectively. 
The set of cuts used for the analysis of the distributions 
corresponds to the angular range $20^\circ$-$160^\circ$. 
Generally, as it can be seen, the level of agreement is quite 
satisfactory. In particular, the electron energy distribution simulated 
by BABAYAGA, well agrees with the exact LABSPV calculation, as 
shown in Fig.~\ref{fig:14}. Because the 
electron energy distribution is driven in the 
present implementation of the PS 
method by the $x$ distribution of the SF, the agreement observed between
exact and PS predictions reinforces,{\it a posteriori}, the interpretation 
of the $x$ variable as residual fraction energy as 
the most natural one in QED PS models. 	
	
Also the PS description of the electron angular variables 
is in nice agreement with the exact calculation, as shown by the 
electron scattering angle and acollinearity distribution of 
Figs.~\ref{fig:15}-\ref{fig:16}. Some disagreement 
is seen in the first acollinearity bins, where 
still acceptable differences at a few per cent level are registered.  
The satisfactory PS predictions for the electron 
angles can be understood since the distributions of these 
variables are dominated, beyond the tree-level approximation, by 
the effect of the longitudinal boost due to the emission of 
unbalanced electron and positron ISR, which, in turn, is 
governed by the $x$ distribution of the electron 
SF.

\section{Conclusions and perspectives}
The LABH process is used at present $\Phi$ and $B$ factories as 
the reference 
reaction to determine the machine luminosity. In order 
to reach a total accuracy at a few 0.1\% level, 
precision calculations of the LABH 
scattering cross section and distributions become more and
more urgent. In particular, the relevant effects due 
to photon emission have to be kept under control and 
accurately simulated in the computational tools required 
by the experimental analysis. 

Along this direction, an original calculation of the 
LABH process has been addressed. It is 
based on a realization of the PS method 
in QED to account for photonic corrections due to 
ISR, FSR and initial-final-state interference. The approach adopted 
allows the calculation of integrated cross sections 
as well as event generation, including
reconstruction of photon $p_\perp$. A new MC event generator 
(BABAYAGA) has been developed and 
is available for a full experimental simulation 
of the LABH process at flavour factories. 
The program has been used to provide several numerical 
results of phenomenological interest, thus showing the 
potential of BABAYAGA in physics analysis. For example, 
it has been shown, both at the level 
of cross section and interesting distributions, 
that the effects due to radiation from 
final-state electrons have to be carefully taken into 
account in the presence of realistic ES at the 
$\Phi$ factories DA$\Phi$NE 
and VEPP-2M, since the impact of FSR 
can be as large as the one due to ISR.  

With the aim of checking the overall (physical+technical) 
precision of the PS approach and relative generator,
a benchmark calculation has been carried out as well. 
It relies upon the exact $O(\alpha)$ calculation 
of the LABH cross section supplemented with  
higher-order LL corrections in the collinear approximation.  
The benchmark computation is also available in the form of
computer code (LABSPV), which is a MC integrator 
allowing for precise cross section calculations, with an 
estimated precision at 0.1\% level. 

By comparing the predictions of the 
PS generator BABAYAGA and of the benchmark calculation 
LABSPV for 
the up to $O(\alpha)$ cross section with typical cuts, it turns out that 
the contribution of the $O(\alpha)$ 
NLO corrections is, as expected, important
in the light of the required theoretical precision, 
being at the 0.5-1\% level for typical 
ES at the $\Phi$ factories. However, it has been also 
shown that the scale $Q^2$ entering in the PS LL calculation 
can be simply adjusted in order to agree within 
0.5\% with the benchmark 
calculation. An energy scale like $Q^2 \approx st/u$ reveals to 
be the best choice that effectively 
reabsorbs $O(\alpha)$ NLO terms into LL 
PS contributions. This conclusion, which holds for 
the integrated cross section, has been proved to be  
generally valid also 
for the most interesting differential distributions. 

With the scale choice $Q^2 = st/u$, the effect of 
higher-order $O(\alpha^n L^n)$ LL corrections has been 
evaluated and found to be at the 1-2\% level, while 
the role of $O(\alpha^2 L)$ corrections 
is marginal, below 0.1\% accuracy. Therefore, one of the 
main conclusions of the present analysis is that, given 
the size of the radiative corrections discussed above,  
theoretical predictions for the LABH process 
at flavour factories aiming at a few 0.1\% 
precision must include the contribution of both  
$O(\alpha)$ NLO terms and
$O(\alpha^n L^n)$ leading logarithmic contributions. 
From the whole of the present analysis, it turns out that 
the present physical precision of BABAYAGA generator is 
0.5\% for typical ES at the $\Phi$ factories. If particularly 
stringent requirements of theoretical accuracy would be in the 
future necessary, the PS approach 
as presently implemented in BABAYAGA should be 
updated by means of an appropriate
merging of the exact $O(\alpha)$ 
matrix element with 
the exclusive photon exponentiation realized by the 
PS algorithm. Some proposals addressing such an issue 
are already available in the literature~\cite{psmm}, in order to 
obtain sensible QCD phenomenological predictions for experiments 
at the TEVATRON and LHC.

A second possible development of the present work concerns 
application of the PS scheme to other phenomenological  
studies of strong interest at $e^+ e^-$ flavour factories. 
In fact, since the PS is a very general method to 
compute photonic radiative corrections, the same formulation here applied to 
the LABH process could be employed to evaluate 
radiative corrections to other large-angle interesting QED processes, 
such as $e^+e^- \to \mu^+\mu^- \, (n\gamma), 
\gamma\gamma \, (n\gamma)$, and also to hadronic final states, 
in particular for  $e^+e^- \to$~hadrons~ and 
$e^+e^- \to$~hadrons~+~$\gamma$, 
the latter of great interest for an energy scan of 
the hadronic cross section below the nominal c.m. energy~\cite{hg}.

A further foreseen development is a phenomenological analysis 
of QED processes at the $B$ factories, along the lines 
followed in the present study.

Such possible developments are by now under consideration.

\section{Acknowledgments}
The authors are very grateful to A.~Bukin, G.~Cabibbo, S.~Eidelman, 
V.N. Ivanchenko, F.~Jegerlehner, V.A.~Khoze, G.~Kukartsev, 
J.~Lee-Franzini, G.~Pan\-ch\-eri, M.~Piccolo, I.~Peruzzi, 
Z.~Silagadze and G.~Venanzoni for useful 
discussions, remarks and interest in their work. 
The authors acknowledge partial support from 
the EEC-TMR Program, Contract N.~CT98-0169.
C.M.~Carloni Calame and C.~Lunardini wish to thank the INFN, Sezione
di Pavia, for the use of computer facilities.


\begin{thebibliography}{99}

\bibitem{dh1}
The DA$\Phi$NE physics handbook, L.~Maiani, G.~Pancheri and N.~Paver eds., 
1992.

\bibitem{dh2}
The second DA$\Phi$NE physics handbook, L.~Maiani, G.~Pancheri and
N.~Paver eds., 1995.

\bibitem{gp} G. Pancheri, Acta Phys. Polonica {\bf 30} (1999) 2243.

\bibitem{cv}
G. Cabibbo and G.~Venanzoni, ``Measuring the DA$\Phi$NE luminosity 
with large
angle Bhabha scattering'', KLOE MEMO n.~168, November 9, 1998.

\bibitem{bmc}
F.A.~Berends and R.~Kleiss, \npb{228}{1983}{537}.

\bibitem{dv}
E.~Drago and G.~Venanzoni, ``A Bhabha generator for DA$\Phi$NE including
  radiative corrections and $\Phi$ resonance'', INFN/AE-97/48, 1997.

\bibitem{priv}
S.~Eidelman and V.N.~Ivanchenko, private communication.

\bibitem{abaetal}
A.B.~Arbuzov et al., \jhep{10}{1997}{001}.

\bibitem{labsmc}
A.B. Arbuzov, ``LABSMC: Monte Carlo event generator 
for large-angle Bhabha scattering'', \hepph{9907298}.

\bibitem{bhwide}
S. Jadach, W. P\l{}aczek and B.F.L. Ward, \plb{390}{1997}{298}.

\bibitem{babar}
See, for example, the home page of the BABAR Event Generator Group {\tt
  http://www.slac.stanford.edu/BFROOT/www/Physics/Tools/generators/}.

\bibitem{fuetal}
J.~Fujimoto et al., \ptp{91}{1994}{333} .

\bibitem{unibab}
H.~Anlauf et al., \cpc{79}{1994}{466}.

\bibitem{lep2bha}
S.~Jadach, O.~Nicrosini et al., ``Event Generators for 
Bhabha Scattering'', in
  Physics at LEP2, G.~Altarelli, T.~Sj\"ostrand and F.~Zwirner eds., CERN
  Report {\bf 96-01}, (CERN, Geneva, 1996), Vol.~2, p.~229.

\bibitem{prd}
See, for example, G.~Montagna, O.~Nicrosini and F.~Piccinini,
  \prd{48}{1993}{1021} and references therein.

\bibitem{gmnp}
M.~Greco, G.~Montagna, O.~Nicrosini and F.~Piccinini, in ref.~\cite{dh2}, 
Vol.~II, p.~629.

\bibitem{vpol}
S.~Eidelman and F.~Jegerlehner, \zpc{67}{1995}{585} .

\bibitem{vpol1}
F.~Jegerlehner, private communication and DESY-99-007, \hepph{9901386}, 
in Proc. of the IVth International Symposium on Radiative Corrections, 
RADCOR98, J. Sola ed., Barcelona, Spain (1998);
J.G.~K\"orner, A.A.~Pivovarov and K.~Schilcher, 
Eur.~Phys.~J {\bf C9} (1999) 551.

\bibitem{ap}
V.N.~Gribov and L.N.~Lipatov, \sjnp{15}{1972}{298}; 
G.Altarelli and G.Parisi,
\npb{126}{1977}{298}; 
Y.L.~Dokshitzer, \jetp{46}{1977}{641}.

\bibitem{rnc}
G.~Montagna, O.~Nicrosini and F.~Piccinini, Riv. Nuovo Cim., Vol.~{\bf
  21} (1998) 1, \hepph{9802302}.

\bibitem{sfcoll}
E.A.~Kuraev and V.S.~Fadin, \sjnp{41}{1985}{466}; 
G.~Altarelli and
G.~Martinelli, in Physics at LEP, J.~Ellis and R.~Peccei eds., CERN 
Report {\bf 86-02} (CERN, Geneva, 1986), Vol.~1, p.~47; 
O.~Nicrosini and
L.~Trentadue, \plb{196}{1987}{551}; \zpc{39}{1988}{479}; 
F.A.~Berends,
G.~Burgers and W.L.~van~Neerven, \npb{297}{1988}{429}; 
M. Skrzypek and
S. Jadach, \zpc{49}{1991}{577}; 
M.~Cacciari et al., Europhys. Lett. {\bf 17} (1992) 123, 
G. Montagna, O. Nicrosini and F. Piccinini, \plb{406}{1997}{243}, 
A.B. Arbuzov, \plb{470}{1999}{252}.

\bibitem{psqcd}
R.~Odorico, \npb{172}{1980}{157}, \plb{102}{1981}{341}; 
P.~Mazzanti and
R.~Odorico, \plb{95}{1980}{133}; 
G.~Marchesini and B.R.~Webber,
\npb{238}{1984}{1}; 
T.~Sj\"ostrand, \plb{310}{1985}{321}; 
K.~Kato and
T.~Munehisa, \prd{39}{1989}{156}; 
K.~Kato, T.~Munehisa and H.~Tanaka,
\zpc{54}{1992}{397}. 
A review of the PS method in QCD and relative QCD
event generators can be found in: R.K.~Ellis, W.J.~Stirling and B.R.~Webber,
QCD and Collider Physics (Cambridge University Press, 1996).

\bibitem{psqed}
J.~Fujimoto, T.~Munehisa and Y.~Shimizu, \ptp{90}{1993}{177}; Y.~Kurihara,
J.~Fujimoto, T.~Munehisa and Y.~Shimizu, \ptp{96}{1996}{1223},
\ibid{95}{1996}{375}.

\bibitem{psohl}
H.~Anlauf et al., \cpc{70}{1992}{97}.

\bibitem{sff}
V.V.~Sudakov, \jetp{3}{1956}{65}.

\bibitem{ape}
G.~Bonneau and F.~Martin, \npb{27}{1971}{381}; V.N.~Baier, V.S.~Fadin and
V.A.~Khoze, \npb{65}{1973}{381}; F.A.~Berends and R.~Kleiss, 
\npb{260}{1985}{32} .

\bibitem{vs}
M.Caffo, E.Remiddi et al., ``Bhabha Scattering'', in $Z$ Physics at LEP1,
G.~Altarelli, R.~Kleiss and C.~Verzegnassi eds., CERN Report {\bf 89-08}
(CERN, Geneva, 1989), Vol.~1, p.171; M.Greco, Riv. Nuovo Cim.
Vol.~{\bf 11} (1988) 1.

\bibitem{hard}
F.A.~Berends et al., \npb{202}{1981}{63} .

\bibitem{stu}
M.~Greco and O.~Nicrosini, \plb{240}{1990}{219}.

\bibitem{a2l}
G.~Montagna, O.~Nicrosini and F.~Piccinini, \plb{385}{1996}{348}.

\bibitem{sabspv}
M.~Cacciari, G.~Montagna, O.~Nicrosini and F.~Piccinini, in Report of the
Working Group on Precision Calculations for the $Z$ Resonance, D.~Bardin,
W.~Hollik, G.~Passarino eds., CERN Report {\bf 95-03} (CERN, Geneva, 1995),
p.~389, \cpc{90}{1995}{301} .

\bibitem{psmm}
F. Krauss, R. Kuhn and G. Soff, 
Acta Phys. Polonica {\bf 30} (1999) 3875; G.~Miu and T.~Sj\"ostrand,
 Phys. Lett. {\bf B449} (1999) 313; G.~Corcella and 
 M.~Seymour, \plb{442}{1998}{417};
 J.~Andr\'e and T.~Sj\"ostrand,
\prd{57}{1998}{5767}; H.~Baer and M.H.~Reno, \prd{44}{1991}{3375}; 
M.H.~Seymour, \cpc{90}{1991}{95}.

\bibitem{hg}
S. Spagnolo, Eur. Phys. J. {\bf C6} (1999) 637;
S. Binner, J.H. Kuhn and K. Melnikov, Phys. Lett. {\bf B459} (1999) 279;
M. Benayoun et al., Mod. Phys. Lett. {\bf A14} (1999) 2605;
A.B. Arbuzov et al., \jhep{12}{1998}{009}.

\end{thebibliography}
\end{document}